\documentstyle[12pt]{article}
\begin{document}
\renewcommand{\baselinestretch}{1.5}

\newcommand\beq{\begin{equation}}
\newcommand\eeq{\end{equation}}
\newcommand\bea{\begin{eqnarray}}
\newcommand\eea{\end{eqnarray}}

\newcommand\Sum{\sum}
\newcommand\Sumi{\sum_i^N}
\newcommand\Sumj{\sum_j^D}
\newcommand\Sumij{\sum_{i,j}^N} 
\newcommand\Sumijk{\sum_{i,j\ne i,k\ne i}^N}
\newcommand\sumijk{\sum_{i\ne j\ne k}^N}
\newcommand\Pij{\prod_{i<j}^N}

\newcommand\expo{e^{-{1\over 2}\Sumi\Sumj x_i^{j2}}}
\newcommand\expon{e^{-{1\over 2}\Sumij x_i^{j2}}}
\newcommand\partialxij{\partial f/\partial x_i^j}
\newcommand\pxijt{{\partial^2 f/\partial x_i^{j2}}}
\newcommand\prl{Phys. Rev. Lett.}
\newcommand\prb{Phys. Rev. {\bf B}}

\centerline{\bf Exactly Solvable Models in Arbitrary Dimensions}
\vskip 1 true cm

\centerline{Ranjan Kumar Ghosh}
\centerline{\it Haldia Government College, P.O. Debhog, }
\centerline{\it Midnapore 721657, India}
\vskip .5 true cm

\centerline{Sumathi Rao \footnote{{\it e-mail
address}: sumathi@mri.ernet.in}} 
\centerline{\it Mehta Research Institute, Chhatnag Road,Jhunsi,}
\centerline{\it Allahabad 221506, India}
\vskip 2 true cm
\noindent {\bf Abstract}
\vskip 1 true cm

We construct  a new class of quasi-exactly solvable many-body Hamiltonians
in arbitrary dimensions, whose ground states can have any correlations
we choose. Some of the known correlations in one dimension and some
recent novel correlations in two and higher dimensions are 
reproduced as special 
cases. As specific interesting examples, we also write down some
new models in two and higher dimensions with novel correlations.
\vskip 1 true cm


\newpage

Exactly solvable models have always attracted a lot of interest
in theoretical physics, because they serve as paradigms for more
complicated realistic models. Examples of such many-body quantum 
Hamiltonians with exact solutions abound in one dimension - \ $e.g.$, 
the Calogero-Sutherland model (CSM) \cite{CSM} and its 
variants \cite {CSM2,CSM3} \ - and they
have proved  useful in the study of 
diverse physical phenomena, such as quantum spin chains
\cite {QSC}, soliton wave propogation\cite{SOLITON} and 
random matrices\cite{RM}.
A particular feature of these models is that their wave-functions
show strong inter-particle correlations. This feature is expected to
be of interest in higher dimensions too, in the study of strongly
correlated systems. Hence, the construction of exactly solvable models
in higher dimensions is of great interest.

A direct generalisation of the CS correlation built into the
wave-function through the Jastrow factor leads to the Laughlin
\cite{LAUGHLIN}
(or Jain\cite{JAIN}) wave-functions in two dimensions, which are 
relevant to the phenomenon of the fractional quantum Hall
effect. Hamiltonians with
ultra-short interactions \cite{TRUGKIV} for
which Laughlin wave-functions are  exact ground state wave-functions
have been known for years.  We have also been able 
to construct  gauge models \cite{US} for which the Laughlin
wave-functions are  exact ground states.

Recently, a new form of the pair correlator was constructed
in two \\ dimensions\cite{DIP1},
which  was anti-symmetric under exchange, but also introduced
zeroes in the wave-function whenever the relative angle between the
position vectors of the two particles was zero or  $\pi$. The Hamiltonian
for which a wave-function constructed from these correlators is
a ground state was also constructed and studied. Later, this
kind of model was generalised to arbitrary dimensions as well
\cite{PIJUSH1}. More recently, Ghosh\cite{PIJUSH2} has constructed
new CS type models with only two-body interactions in arbitrary dimensions.
Although these models, a priori, may appear to have little or no 
physical interest at the present moment, there are two reasons 
for studying such models. One reason is the paucity of exactly
solvable models in higher dimensions, which makes such a study a
worthwhile exercise. Secondly, and more interestingly, it challenges
the ingenuity of the readers to look for realistic systems in 
physics or in other cross-disciplinary areas, where such novel
correlations may be realised.

With this motivation, in this paper, we construct a whole class of exactly
solvable models in arbitrary dimensions. The ground state wave-function
is chosen to be an arbitrary homogeneous function 
(multiplied by an exponential factor) of $DN$ variables
($D$ is the dimension and $N$ is the number of particles), which
is symmetric or anti-symmetric under exchange of the $N$ particles,
depending on whether we are interested in bosonic or fermionic
wave-functions. Then the Hamiltonian is constructed in terms of the
partial derivatives of the wave-function. Further, by writing the Hamiltonian
in terms of appropriate creation and annihilation operators, a class
of excited states and their energies is found.

By taking specific forms of the wave-function, we show that the
CSM and its variants in one dimension as well as the more recent
models in two and arbitrary dimensions are reproduced as
special cases. Interestingly, we find that in two dimensions,
we can construct exactly solvable models for which the more
natural Jastrow-type correlations are relevant. More specifically,
we show that our procedure allows for the construction of a model
for which the unprojected Jain wavefunctions
are exact ground states. As a final example, we construct
a new fermionic Hamiltonian in three dimensions with two and
three body interactions which can be exactly solved for the 
ground state and and a class of excited states. The ground state
wave-function of this model has the novel feature that it vanishes
not only when two particles coincide, but also when their position
vectors differ by multiples of a specified lattice vector
(generalising the result of Murthy et al\cite{DIP1} in two dimensions).


Let us consider a system of $N$ particles in $D$ dimensions
with position vectors ${\bf r}_1 = (x^1_1,x^2_1, \cdots,x^D_1),
{\bf r}_2 = (x^1_2,x^2_2, \cdots,x^D_2), \dots,
{\bf r}_N = (x^1_N,x^2_N, \cdots,x^D_N)$.
We now prove a general mathematical theorem which is applicable 
to this system.
\begin{itemize}
\item{{\bf Theorem I}} : Let $f$ be a function of $ND$ variables  
($i.e.$,  $f \equiv f(x_i^j)$, $i=1,2,\dots,N$, $j=1,2,\dots,D$) 
that is homogeneous and of degree $n$.
Then $\psi = f^{\eta}(x_i^j)\expo$  is\footnote
{Note that $f^{\eta} = g$ is also a homogeneous function
of its variables. Hence, the most general form of the theorem
has $\eta = 1$. However, for future convenience in constructing 
explicit examples, we retain $\eta$. } an exact eigenstate of the
Hamiltonian 
\bea
H = -{1\over 2}\Sumi {\vec \bigtriangledown}_{_i}^2 &+& {1\over 2}
\Sumi {\bf r}_i^2 + {g_1\over 2} \Sumi \Sumj {\pxijt \over f} 
\nonumber\\
&+& {g_2\over 2}  \Sumi \Sumj ({\partialxij \over f})^2
\label{hami}
\eea
when 
\beq
g_1 = \eta \quad {\rm and}\quad  g_2 = \eta(\eta-1)
\label{g1g2}
\eeq
and  with eigenvalue 
\beq
E = {1\over 2}
(2\eta n+ND).
\label{energy}
\eeq
\item{{\bf Proof}} : 
Considered as a function of a single variable $x_i^j$, we see that
\bea
{\partial^2 \psi\over\partial x_i^{j2}} &=& \eta(\eta-1) {(\partialxij)^2
\over f^2} \psi + \eta {\pxijt\over f} \psi \nonumber \\ 
&-& 2\eta x_i^j {\partialxij\over f} \psi - \psi  
+ x_i^{j2} \psi.
\eea 
Summing over all the $ND$ coordinates and using Euler's theorem
for homogeneous functions, 
\beq
\Sumi\Sumj x_i^{j}{\partial f \over \partial x_i^j} \equiv \Sumij 
 x_i^{j}{\partial f \over \partial x_i^j} = nf,
\eeq
we get
\bea
\Sumij 
{\partial^2 \psi\over\partial x_i^{j2}} &=& \eta(\eta-1) 
\Sumij{(\partialxij)^2
\over f^2} \psi  + \eta \Sumij{\pxijt\over f} \psi \nonumber \\    
&-& 2n\eta\psi -ND\psi + \Sumi {\bf r}_i^2 \psi.
\eea
Using this in Eq.(\ref{hami}), we get
\bea
H\psi &=& {1\over 2}[g_2-\eta(\eta-1)]\Sumij {(\partialxij)^2\over f^2}
\psi \nonumber \\ &+& 
{1\over 2}(g_1-\eta)\Sumij{\pxijt\over f} \psi
+{1\over 2} (2n\eta +ND) \psi.
\eea
Hence, for the choice of $g_1$ and $g_2$ given in Eq.(\ref{g1g2}),
$\psi$ is an eigenstate of the Hamiltonian with the eigenvalue 
given in Eq.(\ref{energy}).
\end{itemize}

The next step is to prove that $\psi$ is actually a ground state
wave-function. For this, we formally introduce creation and annihilation
operators as follows - 
\bea
A_x &=&  p_x -ix +i\eta {\partial f/\partial x \over f} \nonumber \\
A_x^{\dagger} &=&  p_x + ix -i\eta {\partial f/\partial x \over f}
\eea
for each of the coordinates $x_i^j$. We then find that the
Hamiltonian in Eq.(\ref{hami}) can be written as 
\beq
H = {1\over 2} \Sumij (A_{x_i^j}^{\dagger} A_{x_i^j})
+ {1\over 2} (2\eta n+ND).
\eeq
$\psi$ is annihilated by all the $A_{x_i^j}$ and is therefore
the ground state of the Hamiltonian with the energy
$(2\eta n+ND)/2$. Note however, that $[A_x, A_y^{\dagger}]$ for
different coordinates $x$ and $y$ do not commute for non-zero
$\eta$ and hence, the Hamiltonian is not fully soluble for the 
excited states.

We can also look for excited states of the Hamiltonian by
making the following ansatz\cite{DIP1} 
for the excited state wave-function - 
\beq
\Phi(x_i^j) = \psi_0(x_i^j) \phi (x_i^j)
\label{ansatz}
\eeq
- $i.e.$, we assume that the excited state wave-function factorises
in terms of the ground state wave-function $\psi_0$ and a completely
symmetric wave-function $\phi$. This form can be justified by looking
at the asymptotic properties of the Hamiltonian.

Using the ansatz in Eq.(\ref{ansatz}) in the original Hamiltonian,
we find that $\phi(x_i^j)$ satisfies 
\beq
-{1\over 2} \Sumij{\partial^2 \phi\over \partial x_i^{j2}}
+\Sumij x_i^j {\partial \phi\over \partial x_i^j}
-\eta \Sumij{\partialxij \over f} {\partial \phi\over \partial x_i^j}
= (E_m-E_0) \phi
\eeq
Here, we have assumed that
\beq
H\Phi = E_m\Phi
\eeq
following \cite{DIP1}.
We now specialise to the case of radial excitations. - 
$i.e.$, we look for solutions where $\phi$ is only a
function of $t=\Sumij x_i^{j2}$. We then find that the
eigenvalue equation for $\phi$ reduces to the confluent
hypergeometric equation given by
\beq
t {d^2\phi\over d t^2} + (E_0-t) {d\phi\over d t} +
{1\over 2}(E_m-E_0)\phi = 0.
\eeq 
The solutions are confluent hypergeometric functions which are
normalisable provided 
\beq
{1\over 2} (E_m-E_0) = m
\eeq
where $m$ is a positive integer. This gives the energy spectrum 
of radially excited states as
\beq
E_m = E_0 + 2m.
\eeq

Note that we have obtained the results for the ground state
wave-function and the excited states in analogy with Ref.\cite{DIP1};
however, our results are much more general and are valid for
{\it any} homogeneous function $f$. In fact, for the function
$\psi = f^{\eta}\expo$ to qualify as an $N$-particle wave-function,
$f^{\eta}$ must  be symmetric/antisymmetric
under exchange for bosons/fermions. This is a requirement that
we need to impose on $f$.
Furthermore, there exists  no
restriction as to dimensionality, number of particles, types of correlation,
or whether there exists two-body, three-body or $N$-body interactions
in the potential. But for useful results, we need to put some
constraints on $f$, such that either the type of correlation present
in the wave-function or the type of interaction present in the 
potential is specified. In either case, we can construct
a whole host of models with exact ground states. But  before we
go on to the construction of new models, let us  first  check that
by choosing appropriate forms of $f$, several earlier known 
models are reproduced as special cases within this formalism.

\begin{itemize}
\item{(I)} One dimensional models (D=1)
\begin{itemize}
\item
a)Let us choose $f$ to be of the form 
\beq
f = \Pij (x_i - x_j)^{\lambda} |x_i-x_j|^{\alpha} \quad 
{\rm and}\quad \psi = f e^{-\Sumi {x_i^2\over 2}}.
\eeq
This is the so-called Jastrow form and is a highly correlated
wave-function. It picks up a phase $(-1)^{\lambda}$ under exchange
of two particles and can be chosen to be fermionic or bosonic
by choosing $\lambda=1$ or $\lambda=0$ respectively.
By differentiating $f$, we can check that
\beq
\Sumi {\partial^2 f/\partial x_i^2\over f} = 
2 \sum_{i<j}^N {(\alpha+\lambda)(\alpha+\lambda-1)\over (x_i-x_j)^2}
\eeq
and the three-body term vanishes, since (without loss of generality),
we may choose $x_1<x_2<\cdots <x_N$ and use the identity
\beq
\sumijk {1\over (x_i-x_j)(x_i-x_k)} = 0.
\label{identity}
\eeq
Thus for this value of $f$, the Hamiltonian in Eq.(\ref{hami}) yields the
Calogero-Sutherland model.

\item
b)A simple  generalisation is to choose $f$ to be of the form
\beq
f=\Pij (x_i^k-x_j^k)^{\lambda}|x_i^k-x_j^k|^{\alpha}
\eeq
and construct the Hamiltonian. ($k=1$ is the CSM.) 
For $k=2$, the three body-term vanishes\cite{PIJUSH2}. But for other values
of $k$, we have both two-body and three-body interactions. 
\end{itemize}

In principle, we can construct many further models with 
multi-particle correlations, since $f$ is arbitrary. For fermionic
wave-functions, $f$ has to be antisymmetric under exchange,
which can always be achieved by choosing $f$ to be of the form
of a determinant multiplied by symmetric factors.

\item{(II)} Two dimensional models (D=2)

Here again, there are several possibilities.
\begin{itemize}
\item {(a)}
As an obvious generalisation of the one dimensional Jastrow
form, we may take  
\bea
f&=&\Pij (|{\bf x}_i|-|{\bf x}_j|)^{\lambda}
||{\bf x}_i|-|{\bf x}_j||^{\alpha}. \nonumber \\
{\rm and} \quad \psi &=& f \expon
\eea
where ${\bf x}_i = (x^1_i,x^2_i)$ and $f$ is antisymmetric under
exchange when $\lambda$ is an odd integer.
The Hamiltonian is easily constructed just as in the one
dimensional case.
The three-body term vanishes by the same logic that it vanishes 
for the one-dimensional case - $i.e.$, since $|{\bf x}_i|$ is a number,
the phase space can be split into identical copies and
within each copy, it is easy to show that the three-body 
interaction term vanishes by the identity in Eq.(\ref{identity}).

\item{(b)}
Recently, a new type of pair correlator was found in two
dimensions with which a Jastrow type correlator could be
constructed. This was of the form $x_i y_j -x_j y_i$.
Using this, one can construct the many-body wave-function
\beq
f = \Pij (x_i y_j -x_j y_i)^{\lambda} 
|x_i y_j -x_j y_i|^{\alpha} 
\eeq
and the corresponding Hamiltonian, which  was 
earlier written down and studied in
Ref.\cite{DIP1}.

\item{(c)}
We may also choose $f$ to be of the form 
\beq 
f = \Pij (|{\bf x}_i|^k - |{\bf x}_j|^k)^{\lambda}.
||{\bf x}_i|^k - |{\bf x}_j|^k|^{\alpha }
\eeq
In this case, using our general procedure, we can write down the
explicit models.  Once again, since $|{\bf x}_i|$ is a number, 
it is easy to check that the three-body term vanishes for $k=2$
(besides $k=1$, of course) and hence for this case, a CS type
model can be constructed.
This model was recently constructed
by Ghosh \cite{PIJUSH2}, 
as  an example of a  CS type model in higher  dimensions.

\end{itemize}

\item{(III)}
Three-dimensional models (D=3) 

\noindent Here, again, some of the lower dimensional models can be
obviously generalised.
\begin{itemize}
\item{(a)}
A direct generalisation of type (c) models in two dimensions yields
the wave-function 
\beq
\psi =  \Pij (|{\bf x}_i|^k - |{\bf x}_j|^k)^{\lambda}.
||{\bf x}_i|^k - |{\bf x}_j|^k|^{\alpha }\expon
\eeq
for which Hamiltonians can be constructed. As before, for 
$k=1$ and $k=2$, three-body terms vanish.

\item{(b)}
A generalisation of the type (b) models has also been earlier 
considered\cite{PIJUSH1}. The simplest way is to consider that the
two vector determinant is now replaced by the three vector determinant,
$i.e.$,
\beq
P_{ij} = \left |
\begin{array}{cc}
 x_i & y_i \\
 x_j & y_j \end{array}
          \right |     \longrightarrow
                          \left |
                  		\begin{array}{ccc}
                                  x_i & y_i & z_i \\
				  x_j & y_j & z_j \\
				  x_k & y_k & z_k \end{array}
			  \right | 
                        = P_{ijk}
\eeq
(${\bf x}_i = (x_i,y_i)$ in two dimensions and ${\bf x}_i = (x_i,y_i,z_i)$ 
in three dimensions). 
$f$ can now be constructed as $f=(P_{ijk})^{\lambda}|P_{ijk}|^{\alpha}$.		

\end{itemize}
 
\end{itemize}
Some models have been considered in even higher dimensions, but we 
shall not consider them explicitly, since our point is merely to
illustrate that all models with polynomial correlations can be
obtained from our general procedure.

Finally, we construct new models in two and three dimensions 
using our general procedure.
\begin{itemize}
\item(i) D=2

\noindent 
Choose $\psi$ to be the unprojected Jain wave-function given by
\beq
\psi = \Pij (z_i-z_j)^{2m} \chi_n e^{-{1\over 2}\Sumi z_iz_i^*}
\eeq
where we have used complex notation and $\chi_n$ is the Slater
determinant of $n$ filled Landau levels - $i.e.$, $\chi_n$
is a determinant involving at most $n-1$ powers of $z_i^*$
in each of its terms. For this Jastrow-Slater form, the
wave-function does not factorise neatly in the usual Jastrow
form. Still by using $f= \Pij (z_i-z_j)^{2m} \chi_n$
and taking the appropriate derivatives as in Eq.\ref{hami},
we can construct the Hamiltonian, which, however, 
has many body interactions and is not simple.

\newpage

\item{(ii)} D=3

\noindent 
Choose $f$ to be of the following form
\bea
f&=& 
\prod_{i<j}^N  
                  \left | \begin{array}{ccc}
                   1 & x_i & x_j \\
                   1 & y_i & y_j \\
                   1 & z_i & z_j  \end{array} \right |^{\lambda}
                   \times {\rm abs} \left |
                   \begin{array}{ccc}
                   1 & x_i & x_j \\
                   1 & y_i & y_j \\
                   1 & z_i & z_j \end{array} \right |^{\alpha}
                    \nonumber\\
&\equiv& \Pij X_{ij}^{\lambda}\times |X_{ij}|^{\alpha} 
     \nonumber \\
    {\rm and} \quad \psi &=& f e^{-{1\over 2} \Sumi (x_i^2+y_i^2+z_i^2)} 
\eea
A fermionic model can always be constructed by choosing $\lambda$
to be an odd integer, since the determinant enforces antisymmetry
under exchange. However, note that the wave-function vanishes not 
only when the relative angle between the two position
vectors is zero or $\pi$ (as in Ref.\cite{DIP1}, 
but whenever the two position vectors
satisfy ${\bf r}_i = m{\bf r}_j + k(1,1,1)$ 
($m,k$ are integers) due to the property of the
determinant.
This wave-function obviously has only two-body correlations 
and can be thought of
as yet another generalisation of case (b) in two dimensions. The
Hamiltonian can be constructed from this wave-function by our general
procedure and is given by
\bea
H &=& -{1\over 2}\Sumi {\vec \bigtriangledown}_{_i}^2 + {1\over 2}
\Sumi {\bf r}_i^2 \nonumber \\
&-&{(\alpha+\lambda)\over 2} \sum_{i,j\ne i} {(y_j-z_j)^2 
+(z_j-x_j)^2 + (x_j-y_j)^2 \over X_{ij}^2}\nonumber  \\
&+&{(\alpha+\lambda)^2\over 2} \Sumi~ \Large[~ \sum_{j\ne i}^N
{(y_j-z_j)\over X_{ij}})^2 + (\sum_{j\ne i}^N
{(z_j-x_j)\over X_{ij}})^2 \nonumber \\
&+& (\sum_{j\ne i}^N {(x_j-y_j)\over X_{ij}})^2 ~~
\Large] .
\eea
This Hamiltonian clearly has  both two-body and three body
interactions. Its energy is given by $E =  ((\alpha + \lambda)
N(N-1) +3N/2)$.

\newpage
\item{(iii)} Generalisations

\noindent
Simple generalisations of this model with two-body correlations  where
\beq 
\left | \begin{array}{ccc}
                   1 & x_i & x_k \\
                   1 & y_i & y_k \\
                   1 & z_i & z_k  \end{array} \right |
                    \longrightarrow
\left | \begin{array}{ccc}
                   1 & x_i^n & x_k^n \\
                   1 & y_i^n & y_k^n \\
                   1 & z_i^n & z_k^n  \end{array} \right |
\eeq
or models with three particle correlations where 
\beq
\left | \begin{array}{ccc}
                   x_i & x_j & x_k \\
                   y_i & y_j & y_k \\
                   z_i & z_j & z_k  \end{array} \right |
                      \longrightarrow
\left | \begin{array}{ccc}
                    x_i^n & x_j^n & x_k^n \\
                    y_i^n & y_j^n & y_k^n \\
                    z_i^n & z_j^n & z_k^n   \end{array} \right |
\eeq
are also possible. 

\end{itemize}

However, our aim is not to write out an exhaustive
set of models. It is more to demonstrate that our general procedure
allows the construction of explicitly soluble Hamiltonians.
For any specified correlation of the ground state wave-function, 
we have a systematic procedure to construct the
Hamiltonian.
And for all these Hamiltonians,  we can construct a class
of excited states based on radial excitations,
using the ansatz in Eq.(\ref{ansatz}).

To conclude, in this letter, we have constructed a class of non-trivial
Hamiltonians, which can be solved exactly for the ground state
and a class of excited states. The ground state can 
have any correlation we choose. In particular, we find that 
any polynomial correlation including the Jastrow and several
recent novel correlations can be reproduced as special
cases. We also construct new exactly solvable fermionic Hamiltonians
in two and three dimensions as an illustration of 
our general procedure.

\section*{Acknowledgments}
We would like to thank Pijush Ghosh for useful discussions.
RKG would like to thank  the Mehta Research Institute
for hospitality during the course of this work.

\end{document}